\documentclass{IEEEtran4PSCC}

\usepackage{cite}

\ifCLASSINFOpdf
   \usepackage[pdftex]{graphicx}
  \DeclareGraphicsExtensions{.pdf,.jpeg,.png}
\else
   \usepackage[dvips]{graphicx}
\fi
\usepackage[cmex10]{amsmath}
\usepackage{amssymb}
\usepackage{mathtools}
\mathtoolsset{centercolon}
\usepackage{bbm}  

\usepackage{array}

\usepackage{todonotes}

\usepackage{multirow}

\DeclareMathOperator*{\argmin}{arg\,min}

\newcommand{\ee}[1]{\mathrm{E}\left[ #1 \right]}

\begin{document}
%
\title{Accelerating System Adequacy Assessment\\using the Multilevel Monte Carlo Approach}


\author{
\IEEEauthorblockN{Simon Tindemans}
\IEEEauthorblockA{Department of Electrical Sustainable Energy \\
Delft University of Technology\\
Delft, The Netherlands\\
s.h.tindemans@tudelft.nl}
\and
\IEEEauthorblockN{Goran Strbac}
\IEEEauthorblockA{Department of Electrical and Electronic Engineering \\
Imperial College London\\
London, UK\\
g.strbac@imperial.ac.uk}
}


\bstctlcite{IEEE:BSTcontrol}

\maketitle

\begin{abstract}
Accurately and efficiently estimating system performance under uncertainty is paramount in power system planning and operation. Monte Carlo simulation is often used for this purpose, but convergence may be slow, especially when detailed models are used. Previously published methods to speed up computations may severely constrain model complexity, limiting their real-world effectiveness. This paper uses the recently proposed Multilevel Monte Carlo (MLMC) framework, which combines outputs from a hierarchy of simulators to boost computational efficiency without sacrificing accuracy. It explains which requirements the MLMC framework imposes on the model hierarchy, and how these naturally occur in power system adequacy assessment problems. Two adequacy assessment examples are studied in detail: a composite system and a system with heterogeneous storage units. An intuitive speed metric is introduced for easy comparison of simulation setups. Depending on the problem and metric of interest, large speedups can be obtained. 
\end{abstract}


\thanksto{
\noindent This research was supported by the SMART-SAFE project, funded through the HubNet (Extension) programme (EPSRC EP/N030028/1).}

\section{Introduction}

Operational and planning problems in the power system domain often involve the assessment of (sub-)system performance across a range of probabilistically modelled scenarios. For all but the simplest power system models, this cannot be done analytically, and Monte Carlo (MC) simulations are used instead. MC simulations are a powerful general purpose computation method with a long tradition in power system applications \cite{Billinton1994}, but convergence to the correct answer may be slow. A number of different \emph{variance reduction} methods exist to speed up convergence of Monte Carlo estimates, e.g. \cite{Billinton1994, Billinton1997a}. One of these, importance sampling, has recently grown in popularity for power system applications, especially in combination with automatic tuning of model bias parameters using the cross-entropy approach \cite{Belmudes2008, DaSilva2010a}. However, implementing importance sampling typically requires deep insight into the model, and limits the design freedom, e.g. for simulations involving complex decision making or sequential actions.  

The Multilevel Monte Carlo (MLMC) method was introduced in the context of computational finance to speed up averaging over sample paths, without compromising model detail or accuracy \cite{Giles2015}. Initial applications involved the combination of multi-resolution models (geometric sequences), but other applications have subsequently evolved. A good overview of the method and its applications is given in \cite{Giles2015}. The MLMC approach has recently been used in a reliability context to speed up the estimation of the average mission time of large systems in \cite{Aslett2017}. In \cite{Huda2017}, electrical distribution system risk metrics were estimated using MLMC, using a multi-scale approach to simulate component failures and repairs. 

This paper considers how the MLMC framework \cite{Giles2015} can be used to accelerate risk calculations, in particular in applications relating to system adequacy assessment of complex systems. The contributions of this work are as follows.
\begin{enumerate}
\item A concise overview of the MLMC approach to the estimation of risks is given. It is shown how the structure required for MLMC simulation naturally occurs in adequacy assessment problems, and can often be implemented with minimal changes to the constituent models. Two examples of common model patterns are given. 
\item An intuitive speed metric is introduced that allows for fair comparison between Monte Carlo simulation approaches, and across risk measures. 
\item Two case studies are presented, each representing one of the common model patterns. The MLMC approach results in large speedups, in one case speeding up simulations by a factor 2000 compared to conventional Monte Carlo sampling. The sensitivity of computational speed to the model stack is investigated.
\end{enumerate}

\section{Methodology}
\subsection{Mathematical problem statement}

Power system performance indicators often take the form of risk measures $q$ that are  expressed as the expectation\footnote{The framework of estimating expectation values is less limiting than it may seem. For example, if one is interested in estimating the \emph{distribution} of $X$ in addition to the expectation $\ee{X}$, one can define a series of quantities $X_{(v)} := \mathcal{I}_{X \le v}$, so that $\ee{X_{(v)}} = F_X(v)$.} of a performance indicator $X$ (a random variable), i.e.  $q = \ee{X}$. Formally, the random variable $X$ may be seen as a function $X: \Omega \rightarrow \mathbb{R}$ that associates a numerical outcome with every system state $\omega \in \Omega$ in a sample space $\Omega$. The probabilistic behaviour of the system, and therefore of $X$, is defined by associating probabilities (sets of) states. 

In the context of system adequacy assessment, the probabilistic behaviour of a power system is typically specified using a bottom up model that defines demand levels, component status, generator output levels, etc. This model generates both the sample space $\Omega$ (the set of all possible combinations of component states) and the associated probabilities. The function $X$ deterministically evaluates any \emph{specific} state $\omega \in \Omega$ and computes a numerical performance measure for that state. The risk measure $q=\ee{X}$ is then the (probability weighted) average of the function $X$ over all states. 

For even moderately complex systems, it is not possible to compute the quantity of interest $q = \ee{X}$ analytically, nor can it be computed by enumeration of all states in $\Omega$. In such cases, it is common to resort to Monte Carlo simulation, in which  power system states $\omega^{(i)}, i=1, 2, \ldots$ are generated using the probabilistic bottom-up model and analysed to provide relevant outcomes $X(\omega^{(i)})$. It should be noted that at any time, multiple outcomes $X_{(a)}, X_{(b)}, \ldots$ (e.g. number of outages, energy not supplied) can be measured simultaneously, at little to no extra cost. In the mathematical analysis that  follows, only a single risk measure $q=\ee{X}$ is discussed, but the methods can trivially be applied in parallel.

\subsection{Conventional Monte Carlo}

A brief summary of conventional Monte Carlo simulation is given in this section, as a point of reference for following sections. In conventional Monte Carlo simulation, the quantity $q=\ee{X}$ is approximated by the Monte Carlo estimator
\begin{equation} \label{eq:MCestimator}
\hat{Q}_{MC} \equiv \frac{1}{n} \sum_{i=1}^n X^{(i)},
\end{equation}
where $\{X^{(1)}, \ldots, X^{(n)}\}$ represents a random sample\footnote{We use the statistics convention that a sample is a set of sampled values, rather than the computational science convention where each $x^{(i)}$ is a sample.} from $X$, with each $X^{(i)}$ independent and identically distributed to $X$. Note that we distinguish the \emph{random variable} $X^{(i)}$ that represents the $i$-th random draw from $X$, and its \emph{realisation} $x^{(i)}$ in a particular experiment or simulation run. The MC estimate for a simulation run is thus given by 
\begin{align}
q \approx \hat{q} &= \frac{1}{n}\sum_{i=1}^n x^{(i)}.
\end{align}

We proceed to use the generic expression \eqref{eq:MCestimator} to reason about the convergence of the result. The error $\Delta Q_{MC}$ obtained in this approximation is 
\begin{equation}
\Delta Q_{MC}= \hat{Q}_{MC} - q.
\end{equation}
The MC estimator $\hat{Q}_{MC}$ is unbiased, and, as a result of the central limit theorem, for a sufficiently large sample size $n$, $\Delta Q_{MC}$ is normally distributed, so that 
\begin{align}
\Delta Q_{MC} & \sim \mathcal{N}\left(0,  \sigma^2_{\hat{Q}_{MC}} \right).
\end{align}
The variance of $\hat{Q}_{MC}$ follows from the MC estimator \eqref{eq:MCestimator}:
\begin{equation}
\sigma^2_{\hat{Q}_{MC}} = \frac{ \sigma^2_X}{n}.\label{eq:varMC}
\end{equation}
As a result, the \emph{standard error} $\sigma_{\hat{Q}_{MC}}$ equals $\sigma_X/\sqrt{n}$, indicating the typical $O(n^{-1/2})$ convergence of MC simulations. 

To quantify the computational efficiency of an MC simulation, we denote by $\tau$ the average time required to generate a single realisation $x^{(i)}$. The time spent to generate a sample of size $n$ is then 
\begin{equation}
t_{MC} = n \tau. \label{eq:timeMC}
\end{equation}
Using this relation, the variance \eqref{eq:varMC} can be expressed as 
\begin{equation}
\sigma^2_{\hat{Q}_{MC}}(t_{MC}) = \frac{ \sigma^2_X \tau}{t_{MC}}. \label{eq:varTimeMC}
\end{equation}

\subsection{Multilevel Monte Carlo}

For multilevel Monte Carlo (MLMC), we assume to have at our disposal a hierarchy of models $\mathcal{M}_1, \ldots, \mathcal{M}_L$ that generate random outputs $X_1,\ldots, X_L$, the expectations of which approximate $\ee{X}$ with increasing accuracy. Specifically, we consider the case where the top level model $\mathcal{M}_L$ is the model of interest, i.e. $q=\ee{X_L}$. The lower level models $\mathcal{M}_0, \ldots, \mathcal{M}_{L-1}$ are approximations of the top level model that are faster to evaluate but have a bias, i.e. $\ee{X_{l<L}} \neq \ee{X_L}$.

The material in this section is generic, and can be found using slightly different notation in e.g. \cite{Giles2015}. The basis for the MLMC method is the trivial identity that is the \emph{telescopic sum}:
\begin{align}
q& =\ee{X_L}\nonumber \\ 
&= \ee{X_0} + \ee{X_1 - X_0} + \ldots + \ee{X_L - X_{L-1}} \nonumber \\
&=r_0 + r_1 + \ldots + r_L.
\end{align}
The quantity of interest $q$ is thus decomposed into a crude estimate $r_0$ plus iterative refinements $r_1, \ldots, r_L$. In MLMC, each of these terms is independently estimated using \eqref{eq:MCestimator}. This results in the MLMC estimator
\begin{subequations}
\begin{align}
\hat{Q}_{ML} & \equiv \sum_{l=0}^L  \hat{r}_l = \sum_{l=0}^L \frac{1}{n_l} \sum_{i=1}^{n_l} Y^{(i)}_l \label{eq:MLMCestimator}\\
\intertext{with}
Y_l^{(i)} &= X_l^{(l,i)} - X_{l-1}^{(l,i)} \label{eq:pairdefinition} \\
X_{-1} & \equiv 0.
\end{align}
\end{subequations}
To clarify notation, for each level $l$ we distinguish the \emph{level outcomes} $X^{(k,i)}_l$, the \emph{level pairs} $(X_l^{(l,i)}, X_{l-1}^{(l,i)})$ and the \emph{level contribution} $\hat{r}_l = (1/n_l)\sum_{i=1}^{n_l} Y^{(i)}_l$.
An additional superscipt has been added to the level outcome $X_l^{(k,i)}$ to denote the level $k$ of the pair it is associated with, because outcomes for level $l$ may be generated differently depending on whether they are paired with outcomes at level $l-1$ or $l+1$, as long as this does not affect the model bias. That is, we require only $\ee{X^{(l,i)}_l} = \ee{X^{(l+1,i)}_l} = \ee{X_l}$. All sampled outcomes are assumed to be mutually independent, \emph{except} for those in a level pair $(X_l^{(l,i)}, X_{l-1}^{(l,i)})$, which are jointly sampled from a common distribution.

The MLMC estimator is unbiased and asymptotically normally distributed, by virtue of the constituent MC estimators of the level contributions. Its variance follows from \eqref{eq:MLMCestimator} and the mutual independence of sampled values:
\begin{align}
\sigma^{2}_{\hat{Q}_{ML}} &= \sum_{l=0}^L \frac{\sigma^2_{Y_l}}{n_l},\label{eq:varMLMC}
\end{align}
\begin{align}
\sigma^2_{Y_l} &= \sigma^2_{X_{l-1}} + \sigma^2_{X_l} -2 \cdot \mathrm{Cov}(X_{l-1}^{(l)}, X_l^{(l)}).
\end{align}
Here, the superscipt $(l)$ on the simulation outputs is maintained, because the covariance term depends on the joint sampling process of the pairs $(X_l^{(l,i)}, X_{l-1}^{(l,i)})$. Clearly, the variance is minimised if the sample pairs are highly correlated.

For a given set of models $\mathcal{M}_0,\ldots, \mathcal{M}_L$, the challenge is to optimally choose the samples sizes $n_l$. Defining the average time to generate a single value $y_l^{(i)}$ as $\tau_l$, the total time taken to produce an MLMC estimate is given by
\begin{equation}
 t_{ML} = \sum_{l=0}^L n_l \tau_l. \label{eq:timeMLMC}
\end{equation}
The optimal sample counts $n_l$ can now be determined by minimising the variance \eqref{eq:varMLMC} with respect to $n_{1:L}$ while keeping $t_{ML}$ constant. Using \eqref{eq:timeMLMC} to substitute $n_0$ and setting $\mathrm{d}\sigma^2_{\hat{Q}_{ML}} / \mathrm{d} n_l = 0$ for $l=1,\ldots, L$ results in optimal sample counts (ignoring their discrete nature)
\begin{equation}
n^*_l = \frac{ t_{ML} }{ \sum_{l'=0}^L \sigma_{Y_{l'}} \sqrt{\tau_{l'}}} \times \frac{\sigma_{Y_l}}{\sqrt{\tau_l}}, \label{eq:optimalCount}
\end{equation}
With this optimal choice of $n_l$, the computational effort spent on each level pair $l$ is proportional to $\sigma_{Y_l} \sqrt{\tau_l}$ (see \eqref{eq:timeMLMC}), and the total variance \eqref{eq:varMLMC} can be expressed as a function of computational time as 
\begin{equation}
\sigma^{*2}_{\hat{Q}_{ML}}(t_{ML}) = \frac{1}{ t_{ML} } \left(  \sum_{l=0}^L \sigma_{Y_{l}} \sqrt{\tau_{l}} \right)^2. \label{eq:optimalvarMLMC}
\end{equation}

\subsection{Measuring simulation speed}

By comparing the expressions for the variance of the conventional and multilevel MC approaches, we can investigate the potential speedup resulting from the MLMC approach. Let us consider the times $\tilde{t}_{MC}$ and $\tilde{t}_{ML}$ required to converge to a given variance $\tilde{v} = \sigma^2_{\hat{Q}_{MC}} (\tilde{t}_{MC})= \sigma^{*2}_{\hat{Q}_{ML}}(\tilde{t}_{ML})$. Then, combining \eqref{eq:varMC} and  \eqref{eq:optimalvarMLMC} results in the expression
\begin{equation}
\textrm{speedup} = \frac{\tilde{t}_{MC}}{\tilde{t}_{ML}} = \left( \frac{\sigma_X \sqrt{\tau}}{ \sum_{l=0}^L \sigma_{Y_{l}} \sqrt{\tau_{l}}} \right)^2. \label{eq:speedup}
\end{equation}
In practice, the variance of the lowest level is similar to that of the direct MC simulator, $\sigma_{Y_0} \approx \sigma_X$, and the cost of evaluating the highest level pair is at least that of a direct evaluation of the highest level, i.e. $\tau_L \ge \tau$. Considerable speedups are possible if $\sigma_{Y_l}\sqrt \tau_l \ll \sigma_X \sqrt \tau$ for all $l$. Intuitively, this occurs when each simplified model $\mathcal{M}_{l-1}$ is much faster than the next level $\mathcal{M}_l$, but returns very similar results for the majority of samples. Examples where this occurs naturally in the context of power system adequacy assessment will be discussed in Sections~\ref{sec:example1} and \ref{sec:example2}.

In order to compare the compuational efficiency of various implementations, we require an operational definition of `computational speed'. Monte Carlo simulations are often run with the goal to estimate the quantity $q$ with a certain relative accuracy, expressed using the coefficient of variation $c_q=\sigma_Q / q$. We note that both \eqref{eq:varTimeMC} and \eqref{eq:optimalvarMLMC} can be brought into the form 
\begin{equation}
\underbrace{\vphantom{ \frac{1}{c_q^2}} \frac{1}{c_q^2}}_{\substack{\text{computational} \\ \text{`distance'} }} 
= \underbrace{ z_q   \vphantom{\frac{1}{c_q^2}} }_\text{speed} \times \underbrace{t  \vphantom{\frac{1}{c_q^2}}}_\text{time}.
\end{equation}
This implicitly defines the computation speed $z_q$ as 
\begin{equation}
z_q := \frac{q^2}{t \sigma^2_{\hat{Q}}(t)}. \label{eq:speeddefinition}
\end{equation}

This definition may be compared with the `figure of merit' used in \cite{Henneaux2014}. The inclusion of the quantity $q^2$ in \eqref{eq:speeddefinition} has a number of advantages, provided that $q\neq 0$. First, the speed has dimensions $1/\mathrm{time}$,  independent of the measure $q$. Second, speeds corresponding to different metrics are directly comparable. For example, when $z_\text{LOLE} < z_\text{EENS}$, this indicates that the LOLE estimator is the limiting factor in achieving convergence to a given coefficient of variation. And finally, the speed metric and the implied computational distance are easily interpretable in terms of simulation outcomes. For example, in order to achieve a coefficient of variation of 1\% (i.e.~a `distance' 10,000) using a speed of $10\,s^{-1}$, a simulation run of $1000\,s$ is required. 

In the course of a simulation run, \eqref{eq:speeddefinition} can be used to estimate the computational speed, replacing $q$ and $\sigma_Q$ by their empirical estimates. The speed $z_q$ for MC and MLMC estimation follow from \eqref{eq:varMC} and \eqref{eq:varMLMC} as 
\begin{align}
z_{q,MC} &= \frac{\hat{q}_{MC}^2}{t_{MC}  \hat{\sigma}^2_X / n}, \\
\hat{\sigma}^2_X &= \frac{\sum_{i=1}^n (x^{(i)} - \frac{1}{n} \sum_{j=1}^n x^{(j)})^2}{n-1},
\intertext{and}
z_{q,ML} &= \frac{\hat{q}_{ML}^2}{t_{ML} \sum_l \hat{\sigma}^2_{Y_l} /n_l }, \\
\hat{\sigma}^2_{Y_l} &= \frac{\sum_{i=1}^{n_l} (y_l^{(i)} - \frac{1}{n_l} \sum_{j=1}^{n_l} y_l^{(j)})^2}{n_l-1}.
\end{align}

\section{Considerations for implementation}

\subsection{Joint sample spaces} 

The core of the MLMC algorithm is the joint generation of sample pairs $(X_l^{(l,i)}, X_{l-1}^{(l,i)})$, used in \eqref{eq:pairdefinition}, in such a way that they are maximally correlated. The random variables $X_l$ and $X_{l-1}$ have sample spaces $\Omega_l$ and $\Omega_{l-1}$, respectively, which must be combined into a joint sample space $\Omega'_l$. We highlight two common model patterns that naturally achieve this.

\subsubsection{Pattern 1: component subsets} \label{sec:approach1}

One common occurrence in system adequacy studies is that the lower level model $\mathcal{M}_{l-1}$ omits components that are present in the higher level model $\mathcal{M}_l$. As a result, the sample space $\Omega_l$ can be written as a Cartesian product
\begin{align}
\Omega_l &= \Omega_{l-1} \times A_l, \label{eq:samplespaces}
\end{align}
where $A_l$ is the sample space of components present in $\mathcal{M}_l$ but not  in $\mathcal{M}_{l-1}$. We may then identify $\Omega'_l$ and $\Omega_l$. In practical terms this means that samples can be generated at the higher level $l$ and unused elements are discarded for the simpler models $\mathcal{M}_{l-1}$. An example of this design pattern is explored in Section~\ref{sec:example1}. 

\subsubsection{Pattern 2: identical randomness} \label{sec:approach2}

It is also easy to conceive of scenarios where $\mathcal{M}_{l}$ and $\mathcal{M}_{l-1}$ have identical sample spaces, so that 
\begin{align}
\Omega'_l = \Omega_l &= \Omega_{l-1}. \label{eq:samplespaces2}
\end{align}
This occurs when both models are driven by the same set of random inputs, but the higher level model performs more complex processing. An example of this model pattern is given in Section~\ref{sec:example2}. 

\subsection{Direct evaluation of expectations}

Occasionally, the base model $\mathcal{M}_0$ is sufficiently simple to permit direct computation of $r_0 = \ee{X_0}$, either analytically or using a numerical approximation procedure. In those cases, the long run efficiency is enhanced by evaluating $r_0$ directly instead of using its MC estimate. The standard deviation $\sigma_{Y_0}$ is then equal to 0, or a value commensurate with the accuracy of the numerical approximation of $r_0$. Although direct evaluation of the lowest level is nearly always preferred, there may be cases where the evaluation of $\ee{X_0}$ is a comparatively time-consuming operation and the optimal trade-off is more complex. In the examples that follow in Sections~\ref{sec:example1} and \ref{sec:example2}, direct evaluation is always possible, and results in faster convergence of the overall MLMC estimator.

The use of an analytical result at the lowest level also highlights a connection between the MLMC method and the control variate approach\cite{Giles2015}. The control variate similarly makes use of a simplified model for which an explicit solution can be calculated. It can therefore be considered as a special case of a bilevel MLMC procedure where the value $\ee{X_0}$ is known and the output $X_0$ is scaled for optimal convergence. The control variate approach was used in \cite{Billinton1997a} to speed up composite system adequacy assessment - a problem that is also addressed  in Section~\ref{sec:example1}.

\subsection{Implementation}

Simulations were implemented in Python~3.7 and were run on an Intel i5-7360U CPU under macOS 10.14.6. A generic multilevel sampler was developed with specialisations for particular simulation studies. No effort was made to optimise the execution speed of individual models, because the aim of this paper is not to maximise execution speed per se, but to investigate the relative speed between sampling strategies. The code used to generate the results in this paper is available \cite{CodeTindemans2020}.

All MLMC simulations started with an exploratory run in which a sample with fixed size $n^{(0)}$ is taken at each level set $Y_l$, in order to determine initial estimates of the evaluation cost $\hat{\tau}_l$ and variance $\hat{\sigma}^2_{Y_l}$. This initial run is followed by a sequence of follow-up runs, each parameterised by a target run time $t^*$. Given $t^*$, optimal sample sizes at each level were determined using \eqref{eq:optimalCount} and the most up to date estimates of evaluation times $\hat{\tau}_l$ and variances $\hat{\sigma}^2_{Y_l}$. For all results in this paper, 10 runs with an estimated run time of 60 seconds (each) were used, for a total run time of approximately 600 seconds. 

One practical concern with determining optimal sample sizes using \eqref{eq:optimalCount} is that the values of $\sigma^2_{Y_l}$ are estimated using relatively small data sets. In power system risk assessment, the simulation outputs $X_l$ often involve measurements of rare events, so that there is a high probability that $Y^{(i)}_l=0$, and therefore $\hat{\sigma}^2_{Y_l}  \ll \sigma^2_{Y_l} $ (or even $\hat{\sigma}^2_{Y_l} =0$). If the estimated value is used naively in \eqref{eq:optimalCount}, this leads to undersampling of $Y_l$, thereby exacerbating the problem because fewer samples are generated that can correct the estimate of $\sigma^2_{Y_l}$. To mitigate this risk, the variance estimators were adjusted as follows. First, a conservative estimate for the variance of $X$ was obtained as
\begin{equation}
\tilde{\sigma}_X^2 = \max_{l} (\hat{\sigma}^2_{X_l}).
\end{equation}
Next, we assumed for the lowest level estimator that $\tilde{\sigma}^2_{Y_0} \approx \tilde{\sigma}_X^2$, and that the ratio $Y_{l+1}/Y_l$ of variances of subsequent level contributions is lower-bounded by a factor $\alpha$. Therefore, updated variance estimates are computed as
\begin{equation}
\tilde{\sigma}^2_{Y_l} = \max(\hat{\sigma}^2_{Y_l}, \alpha^{l} \tilde{\sigma}_X^2 ),
\end{equation}
for those pairs $l$ where $\ee{Y_l}$ is estimated by sampling. For the simulations, the value of $\alpha$ was heuristically set to 0.1. 

Finally, in simulations, multiple risk measures $q_{(a)}, q_{(b)}, \ldots $ were estimated in parallel. In determining optimal sample sizes, one of these was selected as the `target measure' to optimise for, so that its mean and variance estimates were inserted in \eqref{eq:optimalCount} to determine the optimal allocation of sample counts $n_l$.

\section{Composite system adequacy assessment}\label{sec:example1}

The first case study is a system adequacy assessment of the single area IEEE Reliability Test System (RTS) \cite{Grigg1999}. A two-level MLMC approach is used, where the upper level, i.e. the study of interest, is a hierarchical level 2 (HL2) study \cite{Billinton1994}: a composite system adequacy assessment that takes into account transmission line outages and constraints. The lower level HL1 is a single node assessment that omits the transmission system. This is in accordance with the subset model pattern in Section~\ref{sec:approach1}.

\subsection{Models}

\subsubsection{Model $\mathcal{M}_1$ - Composite system adequacy assessment (HL2)}
The RTS model defines outage probabilities of generators and transmission lines, which were modelled as independent two state Markov models. Maintenance and transient outages were not considered. Load levels were sampled by uniformly selecting an hour from the annual demand trace and assigning loads to each node in proportion to the maximum nodal demands. 

Therefore, at the \emph{upper level} ($l=1$), a sampled system state $\omega^{(i)}_1$ consists of: (i) the nodal demand $d^{(i)}_n$ for $n \in \mathcal{N}$, the set of nodes; (ii) the generator status $\gamma^{(i)}_j \in \{0,1\}$ for $j \in \mathcal{G}$, with $\mathcal{G} = \bigcup_{n \in \mathcal{N}} \mathcal{G}_n$, where $\mathcal{G}_n$ is the set of generators in node $n$; (iii) the line status $\lambda^{(i)}_k \in \{0,1\}$ for $k \in \mathcal{L}$, the set of transmission lines. Let generator and line flow limits be given by $g_j^{\mathrm{max}}$ and $f_k^{\mathrm{max}}$. Then, the amount of curtailment $C_2$ is computed by the linear program
\begin{align} \label{eq:hl2}
& C_2(\omega^{(i)}_1) = \min_{c_{1:|\mathcal{N}|}, g_{1:|\mathcal{G}|}} \sum_{n \in \mathcal{N}} c_n,
\end{align}
subject to
\begin{align}
 0 \le & c_n  \le d_n^{(i)}, & \forall n \in \mathcal{N}& \nonumber\\
  0 \le & g_j  \le \gamma_j^{(i)} g^{\mathrm{max}}_j,  & \forall j  \in \mathcal{G} &\nonumber \\
-f^{\mathrm{max}}_k \le & \sum_{n \in \mathcal{N}}  M^{(i)}_{k n} [ \sum_{j \in \mathcal{G}_n} g_j + c_n - d_n^{(i)} ]  \le  f^{\mathrm{max}}_k, & \forall k  \in \mathcal{L}  & \nonumber\\
0=& \sum_{n\in \mathcal{N}} [\sum_{j \in \mathcal{G}_n} g_j + c_n - d_n^{(i)}]  ,&&\nonumber
\end{align}
where the matrix $M^{(i)}=D^{(i)} A (A^T D^{(i)}  A + 1/|\mathcal{N}|)^{-1}$ relates bus injections and line flows. The directed line-node incidence matrix $A$ has elements $+1$ for outgoing lines and $-1$ for incoming lines; the diagonal matrix $D^{(i)}$ has elements $D^{(i)}_{kk}= \lambda_k^{(i)}/x_k$, where $x_k$ is the reactance of line $k$. The element-wise constant $1/|\mathcal{N}|$ ensures invertibility, eliminating the need for a designated slack bus. In cases where line outages resulted in multiple islands, problem~\eqref{eq:hl2} was formulated and solved for each island independently and the curtailments were summed to obtain the total system curtailment. Linear optimisation was performed using \texttt{scipy.optimize.linprog}, with the revised simplex method.

\subsubsection{Model $\mathcal{M}_0$ - Generation adequacy assessment (HL1)}
For HL1 assessment, a single-node generation adequacy analysis is performed, without transmission line constraints and outages. The \emph{lower level} system state $\omega^{(i)}_0$ can thus be obtained from $\omega^{(i)}_1$ by omitting the line status variables. For this HL1 study, the curtailment is calculated as
\begin{equation} \label{eq:hl1}
C_1(\omega_0^{(i)}) = \max \left( 0, \sum_{n \in \mathcal{N}}\left[ d_n^{(i)} - \sum_{j \in \mathcal{G}_n} \gamma_j^{(i)} g^{\mathrm{max}}_j  \right] \right).
\end{equation}

\subsubsection{Risk measures}
Two common risk measures were computed: the probability of load curtailment (PLC) and expected power not supplied (EPNS). The related performance measures $X_{q,l}$ are defined in terms of the load curtailment \eqref{eq:hl1} and \eqref{eq:hl2} as
\begin{align}
X_{\mathrm{PLC},l}(\omega) &= \mathbbm{1}_{C_l(\omega)>0}, \label{eq:LOLoutput} \\ X_{\mathrm{EPNS},l}(\omega) &= \max(0,C_l(\omega)).
\end{align}

\begin{table}[!t]
\renewcommand{\arraystretch}{1.3}
\centering
\caption{Composite system adequacy assessment - available models}
\label{tab:singlemodelresults1}
\begin{tabular}{|c|c|c|c|c|}
\hline
model & description & $z_{\mathrm{PLC}}$ [1/s] & $z_{\mathrm{EPNS}}$ [1/s] & direct evaluation \\ 
\hline
$\mathcal{M}_1 $ & HL2& $0.31$ & $0.17$  & no  \\
$\mathcal{M}_0 $ & HL1 & $34.3^*$ & $18.6^*$  & optional \\
\hline
\end{tabular}\\
\vspace{0.1cm}
${}^*$: inherent estimation bias
\end{table}

\begin{table*}[!h]
\renewcommand{\arraystretch}{1.3}
\centering
\caption{Composite system adequacy assessment - comparison of approaches}
\label{tab:results1}
\begin{tabular}{|c|c|c|c|c|c|c|c|c|}
\hline
& & & \multicolumn{3}{c|}{PLC estimation}  & \multicolumn{3}{c|}{EPNS estimation}  \\
estimator & models used & run time [s] & \multicolumn{1}{c}{PLC}  & \multicolumn{1}{c}{$z_{\mathrm{PLC}}$ [1/s]} & speedup & \multicolumn{1}{c}{EPNS [MW]}  & \multicolumn{1}{c}{$z_{\mathrm{EPNS}}$ [1/s]} & speedup \\
\hline
MC & $\mathcal{M}_1$ & 582 & $1.71(13)\times 10^{-3}$ & 0.31 & n/a & 0.238(24) & 0.17 & n/a \\
MLMC (sampling) & $\mathcal{M}_1$, $\mathcal{M}_0$ & 627 & $1.50(7)\times 10^{-3}$ & 0.79 & 2.5 & $  0.190(6)$ & 1.73 & 10\\
MLMC (with expectation) & $\mathcal{M}_1$, $\mathcal{M}_0$ &  601 & $1.48(6)\times 10^{-3}$ & 1.04 & 3.3 & $0.186(5)$ & 2.54 & 15\\
\hline
\end{tabular}
\end{table*}

\subsection{Results}

Throughout, Monte Carlo estimates of risk measures are given with the relevant number of significant digits, followed by the estimated standard error in parentheses. Thus, $1.71(13)\times 10^{-3}$ stands for an estimate of $0.00171$ with a standard error of $0.00013$. For all MLMC runs, an initial exploratory run with $n^{(0)}=100$ was used, followed by 10 runs of approximately 60 seconds. The target risk measure for sample size optimisation was EPNS. Unless stated otherwise, thermal line ratings were scaled to 80\% of the nominal values, to tighten network constraints. 

Table~\ref{tab:singlemodelresults1} compares the speed obtained with the individual models for the estimation of both PLC and EPNS risk measures, and whether direct evaluation of the expectation is possible (for an effective `sampling speed' of $z=\infty$). These numbers were estimated at the end of 10-minute conventional MC runs. The HL1 model is over one hundred times faster than the HL2 model for both measures, but of course this comes at the cost of an estimation bias. 

Table~\ref{tab:results1} compares the results of three different estimators. The top row is the conventional Monte Carlo estimator that directly performs the HL2 study. The middle row represents a two-level MLMC approach where HL2 sampling is combined with HL1 sampling, immediate leading to significant speedups of 2.5 (for PLC) and 10 (for EPNS). In the third configuration (bottom row), further speedups are obtained by eliminating sampling of the lower level model, and computing the lower level estimates $r_{\mathrm{PLC},0}=\ee{X_{\mathrm{PLC},0}}$ and  $r_{\mathrm{EPNS},0}=\ee{X_{\mathrm{EPNS},0}}$ directly by convolution using 1~MW discretisation steps. 

An interesting observation is that, for the regular MC sampler, the speed of PLC estimation (0.31 s$^{-1}$) is larger than that of EPNS estimation (0.17 s$^{-1}$). However, the MLMC sampler sees much more substantial speedups for EPNS estimation than for PLC estimation. This is only partially caused by the EPNS-focused sample size optimisation. The other factor is that the discontinuous loss-of-load indicator \eqref{eq:LOLoutput} is less amenable to successive approximation \cite{Giles2015}.

\begin{table}[!bt]
\renewcommand{\arraystretch}{1.3}
\centering
\caption{Composite system adequacy assessment - multilevel contributions}
\label{tab:study1breakdown}
\begin{tabular}{|c|c|c|c|c|}
\hline
term &  PLC & EPNS [MW] & $\tau_l$ [ms] & $n_l$ \\
\hline
$\hat{r}_1$ & $4.0(7)\times 10^{-4}$ & $0.051(5)$ & $5.4$ & 93\,158 \\
$\hat{r}_0$   & $1.101(16)\times 10^{-3}$ & $0.139(3)$ &$0.023$ & 4\,380\,194  \\
\hline
sum & $1.50(7)\times 10^{-3}$ & $0.190(6) $& \multicolumn{1}{c}{} &\multicolumn{1}{c}{}  \\
\cline{1-3}
\end{tabular}
\end{table}

Table~\ref{tab:study1breakdown} gives insight into the multilevel structure of the regular MLMC estimate. For both PLC and EPNS, the refinement term $\hat{r}_1$ is substantially smaller than the crude estimate $\hat{r}_0$. More importantly, sampling from the HL1 model is substantially faster (0.023 ms per evaluation) than the HL2-HL1 difference term (5.4 ms per evaluation), due to the linear program~\eqref{eq:hl2} involved in the latter. The MLMC algorithm adapts to this cost difference by invoking the HL1 model nearly 50 times as often.

Finally, Table~\ref{tab:study1constraints} shows the impact on convergence speed of varying the thermal line ratings between 80\% and 100\% of the nominal values. Higher line ratings cause fewer constraints, which results in a slight reduction in speed for the regular MC sampler. On the other hand, the MLMC sampler experiences very large speedups as the difference between the results from the HL1 and HL2 models becomes smaller, so that fewer (expensive) HL2 evaluations are required. Once again, the gains in EPNS estimation speed exceed the gains in PLC estimation speed.

\begin{table}[!ht]
\renewcommand{\arraystretch}{1.3}
\centering
\caption{Composite system adequacy assessment - thermal ratings}
\label{tab:study1constraints}
\begin{tabular}{|c|c|c|c|c|c|c|}
\hline
relative &  \multicolumn{3}{c|}{PLC estimation}  & \multicolumn{3}{c|}{EPNS estimation}  \\
line rating &  \multicolumn{1}{c}{$z^{MC}$ }  & \multicolumn{1}{c}{$z^{ML}$} & speedup & \multicolumn{1}{c}{$z^{MC}$ }  & \multicolumn{1}{c}{$z^{ML}$ } & speedup  \\
\hline
0.8 & 0.31& 1.04& 3.3 &0.17 & 2.54 & 15 \\
0.9 & 0.26 & 1.38 & 5.3 & 0.14 & 4.69 & 34 \\
1.0 & 0.25 & 2.11 & 8.6 &0.12 & 16.7 & 143  \\
\hline
\end{tabular}
\end{table}

\section{Dispatch of storage}\label{sec:example2}

The second example concerns the assessment of system adequacy in the presence of energy-constrained storage units (e.g. batteries). The energy constraints couple decisions in subsequent time slots, thus necessitating the use of time-sequential Monte Carlo simulations. Convergence for time-sequential simulations tends to be much slower than for snapshot problems, due to significant correlations in visited system states. An additional complication is deciding an appropriate dispatch strategy  for energy storage units. A greedy EENS-minimising discharging strategy was recently proposed in \cite{Evans2019}, as a reasonable default dispatch strategy for adequacy studies. 

\subsection{Models}

The Great Britain (GB) adequacy study from \cite{Evans2019} is reproduced here, with an eye on speeding up estimation of loss of load expectation (LOLE) and expected energy not supplied (EENS) risks using the MLMC approach.  Individual simulations are run for a sequence of 8760 hours (1 year). The system performance in a simulated year is driven entirely by the net generation margin trace
\begin{equation}
M_t(\omega^{(i)}) = g^{(i)}_t +w^{(i)}_t - d^{(i)}_t, \quad t \in \{1,\ldots, 8760\},
\end{equation}
where the sampled state $\omega^{(i)}$ consists of the demand trace $d_t^{(i)}$, wind power trace $w_t^{(i)}$ and conventional generation trace $g_t^{(i)}$. Annual demand traces are chosen randomly from historical GB demand measurements for 2006-2015 (net demand, \cite{Staffell2018}). Annual wind traces are similarly sampled from a synthetic data set for hypothetical GB wind power output for the period 1985-2014, derived from MERRA reanalysis data and an assumed constant distribution of wind generation sites with an installed capacity of 10~GW \cite{Staffell2016}. Conventional generation traces are generated using an assumed diverse portfolio of thermal units. The portfolio of 27 storage units was based on storage units contracted in the GB 2018 T-4 capacity auction. The reader is referred to \cite{Evans2019} for further model details.

We consider four different storage dispatch models. The resulting storage dispatch (with sign convention that consumption is positive) is denoted by $S_{t,l}(\omega)$, and is entirely determined by the net generation margin $M_t(\omega^{(i)})$. All four models are defined on the same sample space $\Omega$, providing an example of the model pattern described in Section~\ref{sec:approach2}. However, the models differ tremendously in computational complexity, as is clear from the descriptions below.

\begin{table}[!b]
\renewcommand{\arraystretch}{1.3}
\centering
\caption{Time-sequential simulation with storage - available models}
\label{tab:singlemodelresults2}
\begin{tabular}{|c|c|c|c|c|}
\hline
 &  &  &  & direct  \\ 
model & description & $z_{\mathrm{LOLE}}$ [1/s] & $z_{\mathrm{EENS}}$ [1/s] &  evaluation \\ 
\hline
$\mathcal{M}_2 $ & EENS-optimal & $0.105$ & $0.053$  & no \\
$\mathcal{M}_1 $ & sequential  & $0.83^*$ & $0.38^*$  & no \\
$\mathcal{M}_0 $ & average  & $41.9^*$ & $22.1^*$  & optional  \\
$\mathcal{M}'_0 $ & no storage & $61.7^*$ & $34.8^*$  & optional  \\
\hline
\end{tabular}\\
\vspace{0.1cm}
${}^*$: inherent estimation bias
\end{table}

\begin{table*}[!h]
\renewcommand{\arraystretch}{1.3}
\centering
\caption{Time-sequential simulation with storage - model comparison}
\label{tab:study2results}
\begin{tabular}{|c|c|c|c|c|c|c|c|c|}
\hline
 & &run & \multicolumn{3}{c|}{LOLE estimation}  & \multicolumn{3}{c|}{EENS estimation}  \\
estimator &models used& time [s] & \multicolumn{1}{c}{LOLE [h/y]}  & \multicolumn{1}{c}{$z_{\mathrm{LOLE}}$ [1/s]} & speedup & \multicolumn{1}{c}{EENS [MWh/y]}  & \multicolumn{1}{c}{$z_{\mathrm{EENS}}$ [1/s]} & speedup \\
\hline
MC & $\mathcal{M}_2$ & 620 & $1.54(19)$ & 0.105 & n/a & 2\,100(400) & 0.053 & n/a \\
MLMC (3 layer with no-store) & $\mathcal{M}_2$,$\mathcal{M}_1$,$\mathcal{M}'_0$ & 636 & $1.59(6)$ & 1.10 & 10 & $  2\,275(71)$ & 1.61 & 30 \\
MLMC (2 layer with average) &$\mathcal{M}_2$,$\mathcal{M}_0$  & 618 & $1.75(5)$ & 1.88 & 18 & $  2\,415(16)$ & 38.1 & 719 \\
MLMC (3 layer with average) & $\mathcal{M}_2$,$\mathcal{M}_1$,$\mathcal{M}_0$  & 615 & $1.72(3)$ & 6.88 & 66 & $2\,397(9)$ & 112 & 2\,113\\
\hline
\end{tabular}
\end{table*}

\subsubsection{Model $\mathcal{M}_2$ - EENS-optimal dispatch}
The storage dispatch $S_{t,2}(\omega)$ is computed using the EENS-minimising algorithm given in \cite{Evans2019}. It is sequential and requires complex logic for each step.

\subsubsection{Model $\mathcal{M}_1$ - Sequential greedy dispatch}
The storage dispatch $S_{t,1}(\omega)$ is computed using a heuristic approximation of the EENS-minimising policy. Storage units $s \in \mathcal{S}$ are sorted by decreasing \emph{time to go} (from full) $\overline{e}_s/\overline{p}_s$, where $\overline{e}_s$ and $\overline{p}_s$ are energy and discharge power ratings, respectively. Then, a sequential greedy dispatch is performed, charging when possible, and discharging only when required to avoid load curtailment. Evaluating this model requires one sequential pass per storage unit, but the simulation steps are trivial.

\subsubsection{Model $\mathcal{M}_0$ - Constant peak-shaving dispatch}
The storage fleet is optimistically approximated by a single storage unit with $\overline{e} = \sum_s \overline{e}_s$ and $\overline{p} =\sum_s \overline{p}_s$. A mean daily demand profile $\tilde{d}_{1:24}$ is computed by averaging demand over all historical days. This profile is used to compute a \emph{single} daily dispatch pattern $\tilde{s}_{1:24}$ that solves to following quadratic optimisation problem to flatten the average total demand profile $(\tilde{d}_h+\tilde{s}_h)_{h=1:24}$:
\begin{equation}
\tilde{s}_{1:24}  = \argmin_{s_{1:24},e_{1:24}} \sum_{h=1}^{24} (\tilde{d}_h + s_h)^2, \\
\end{equation}
subject to
\begin{align}
-\overline{p} & \le s_h \le \overline{p}, & h=1,\ldots, 24 &  \nonumber \\
0 & \le e_h \le \overline{e}, & h=1,\ldots, 24 & \nonumber \\
e_{h+1}  & = e_h + s_h \times 1~\mathrm{ hour} ,& h=1,\ldots, 23 & \nonumber \\
 e_1 & = e_{24} + s_{24} \times 1~\mathrm{ hour}. \nonumber
\end{align}
This problem was solved using the Python \texttt{quadprog} package. The resulting annual storage dispatch is obtained by repeating the 24-hour dispatch pattern:
\begin{equation}
S_{t,0}(\omega) = \tilde{s}_{(t \bmod 24)}.
\end{equation}
Because $S_{t,0}$ is a deterministic load offset, risk measures for this model can be computed by convolution.  

\subsubsection{Model $\mathcal{M}'_0$ - No storage} 
This alternative lowest level model does not use storage at all, so that $S_{t,0}=0$.

\subsubsection{Risk measures}
The net generation margin $M_t(\omega)$ and storage dispatch $S_{t,l}(\omega)$ result in a curtailment trace as follows
\begin{equation}
C_{t,l}(\omega) = \max[0, - M_t{(\omega)} + S_{t, l}(\omega)], \quad \forall t.
\end{equation}
The LOLE and EENS risk measures can be computed using the performance measures
\begin{align}
X_{\mathrm{LOLE}, l}(\omega) &= \sum_{t=1}^{8760} \mathbbm{1}_{C_{t,l}(\omega) >0}, \\
X_{\mathrm{EENS}, l}(\omega) &= \sum_{t=1}^{8760} C_{t,l}(\omega) \times 1h.
\end{align}

\begin{table}[b]
\renewcommand{\arraystretch}{1.3}
\centering
\caption{Time-sequential simulation with storage - multilevel contributions}
\label{tab:study2breakdown}
\begin{tabular}{|c|c|c|c|c|}
\hline
term &  LOLE [h/y] & EENS [MWh/y] & $\tau_l$ [ms] & $n_l$ \\
\hline
$\hat{r}_2$ & $0(0)$ & $-0.6(4)$ & $1\,670$ & 190  \\
$\hat{r}_1$ & $-0.42(3)$ & $-150(9)$ & $167$ & 1\,771 \\
$\hat{r}_0$   & 2.14 & $2\,548$ & n/a & n/a \\
\hline
sum & $1.72(3)$ & $2\,397(9) $& \multicolumn{1}{c}{} &\multicolumn{1}{c}{}  \\
\cline{1-3}
\end{tabular}
\end{table}

\subsection{Results}

Table~\ref{tab:singlemodelresults2} compares the speed obtained with the individual models for the estimation of both LOLE and EENS risk measures, and whether direct evaluation of the expectation is possible with each model. All numbers were estimated at the end of 10-minute conventional MC runs. Very large differences in model speed are visible, with the detailed model $\mathcal{M}_2$ being over 500 times slower than the crude model $\mathcal{M}'_0$. 

For MLMC simulations, an exploratory run with $n^{(0)}=20$ was used, followed by 10 runs of 60 seconds, where sample sizes were optimised for the EENS risk measure. In all cases, the crude estimate $r_0=\ee{Y_0}$ was evaluated using a convolution approach.  Results are shown in Table~\ref{tab:study2results}, comparing the performance of three MLMC architectures with direct MC simulation. A three-layer architecture using model $\mathcal{M}'_0$ (without storage) as a bottom layer achieved speedups of 10 (LOLE) and 30 (EENS), but much better results were obtained when the daily average dispatch model $\mathcal{M}_0$ was used - even when a two-layer MLMC stack was created by omitting the intermediate sequential greedy dispatch model. 

The results show that the MLMC performance is very sensitive to the choice of levels, but robust speedups are available even for sub-optimal model choices. The best performing architecture is further analysed in Table~\ref{tab:study2breakdown}. It can be seen that the contribution from the final refinement $\hat{r}_2$ is minimal, i.e. the heuristic model is very accurate, which is key to the observed speedup of $2113$. The MLMC algorithm dynamically adjusted sample sizes to generate more samples evaluating $Y_1=X_1-X_0$ than on the costly evaluation of $Y_2=X_2-X_1$. Moreover, no samples are spent on the contribution $\hat{r}_0$, which can be computed directly by convolution. As a result, the speed $z_{\mathrm{EENS}}$ is able to exceed even that of the fastest model in Table~\ref{tab:singlemodelresults2} (for regular MC estimation). 

\section{Conclusions and future work}
This paper has set out how the MLMC approach can be applied to power system risk analysis, and specifically to system adequacy assessment problems. Common model patterns were identified that are particularly amenable to MLMC implementation, and a computational speed measure \eqref{eq:speeddefinition} was introduced to quantify simulation speed in a way that is easily comparable across tools, Monte Carlo methods and risk measures. Two case studies illustrate the potential for speeding up estimation of risk measures, and the ability to apply the method to complex simulations. 

In future work, we will consider automatic selection of optimal model stacks, and explore the scope for the application of multi-index Monte Carlo schemes\cite{Haji-Ali2016} . 

\section*{Acknowledgments}

We thank Kate Ward and Iain Staffell for the provision of GB demand and wind power data, and Michael Evans for sharing Python code for the EENS-minimising dispatch. We are also grateful for insightful discussions with Chris Dent, Matthias Troffaes, Amy Wilson and Stan Zachary.

\bibliographystyle{IEEEtran}

%

\end{document}